\pdfoutput=1
\documentclass{article}

\usepackage[final]{neurips_2022_ml4ps}
\usepackage{natbib}

\newcommand{\deeplenstronomy}{\texttt{deeplenstronomy}}
\newcommand{\sbi}{\texttt{sbi}}
\newcommand{\lenstronomy}{\texttt{lenstronomy}}
\newcommand{\einsteinrad}{$\theta_{ein}$}

\usepackage[utf8]{inputenc} 
\usepackage[T1]{fontenc}    
\usepackage{hyperref}       
\usepackage{url}            
\usepackage{booktabs}       
\usepackage{amsfonts}       
\usepackage{nicefrac}       
\usepackage{microtype}      
\usepackage{xcolor}         
\usepackage{graphicx}

\usepackage{amsmath}
\usepackage{amssymb}
\usepackage{mathtools}
\usepackage{amsthm}

\title{Strong Lensing Parameter Estimation on Ground-Based Imaging Data Using Simulation-Based Inference}

\author{%
  Jason Poh \\
 Department of Astronomy and Astrophysics,\\
The University of Chicago\\
  \texttt{jasonpoh@uchicago.edu} \\
   \And
Ashwin Samudre \\
  School of Computing Science,\\
   Simon Fraser University\\
  \texttt{ashwin\_samudre@sfu.ca} \\
   \And
  Aleksandra \'Ciprijanovi\'c \\
  Fermi National Accelerator Laboratory\\
  \texttt{aleksand@fnal.gov} \\
   \And
   Brian Nord \\
  Fermi National Accelerator Laboratory;\\
 Kavli Institute for Cosmological Physics \&\\ 
Department of Astronomy and Astrophysics,\\
The University of Chicago; \\
Laboratory for Nuclear Physics, MIT \\
   \texttt{nord@fnal.gov} \\
\And
   Gourav Khullar \\
  Department of Physics and Astronomy;\\
PITT PACC,\\ 
University of Pittsburgh\\
   \texttt{gourav.khullar@pitt.edu} \\
\And
   Dimitrios Tanoglidis \\
  Department of Physics and Astronomy;\\
  Data Driven Discovery Initiative,\\
University of Pennsylvania\\ 
   \texttt{dtanogli@sas.upenn.edu} \\
\And
   Joshua A. Frieman \\
    Department of Astronomy and Astrophysics \&\\
    Kavli Institute for Cosmological Physics,\\
   The University of Chicago;\\
 Fermi National Accelerator Laboratory \\
   \texttt{jfrieman@uchicago.edu} \\
}

\begin{document}

\vspace{-20pt}
\maketitle

\begin{abstract}
\vspace{-5pt}

Current ground-based cosmological surveys, such as the Dark Energy Survey (DES), are predicted to discover thousands of galaxy-scale strong lenses, while future surveys, such as the Vera Rubin Observatory Legacy Survey of Space and Time (LSST) will increase that number by 1-2 orders of magnitude. 
The large number of strong lenses discoverable in future surveys will make strong lensing a highly competitive and complementary cosmic probe.

To leverage the increased statistical power of the lenses that will be discovered through upcoming surveys, automated lens analysis techniques are necessary. We present two Simulation-Based Inference (SBI) approaches for lens parameter estimation of galaxy-galaxy lenses. We demonstrate the successful application of Neural Posterior Estimation (NPE) to automate the inference of a 12-parameter lens mass model for DES-like ground-based imaging data. We compare our NPE constraints to a Bayesian Neural Network (BNN) and find that it outperforms the BNN, producing posterior distributions that are for the most part both more accurate and more precise; in particular, several source-light model parameters are systematically biased in the BNN implementation.

\end{abstract}
\vspace{-5pt}
\section{Introduction}
\vspace{-2mm}
Strong gravitational lensing systems are well-established observational probes of dark matter and dark energy. 
On the astrophysical side,
the morphology of lensed images provides information about the distribution of dark and baryonic matter in lens systems \citep[e.g.,][]{Auger2010a,Barnabe2011,Newman2015,Newman2012,Newman2012a}. 
On the cosmological front, strong lensing systems with multiple images of a time-varying source (e.g. quasars and supernovae) can be used to constrain the Hubble constant as well as models of dark energy \citep[e.g.,][]{Suyu2010,Suyu2013,Shajib2020}.


Approximately a thousand strong lensing systems have been discovered to date. 
Wide-field surveys provide prime opportunities to discover more strong lensing systems for future follow-up observations. 
The number of observed lensed systems will increase with the current state-of-the-art and with future large-scale astronomical surveys, some of which are ground-based, such as the Dark Energy Survey~\citep[DES;][]{DES2016}, the Hyper Suprime-Cam Subaru Strategic Program~\citep[HSC-SSP;][]{HSC2018}, the Vera Rubin Observatory Legacy Survey of Space and Time~\citep[LSST;][]{IK2019}, and some which are space-based, such as Euclid\footnote{\url{https://www.cosmos.esa.int/web/euclid}} and the Nancy Grace Roman Space Telescope\footnote{\url{https://roman.gsfc.nasa.gov}}.
The number of lenses that will be discovered through upcoming surveys will make lens modeling through conventional techniques extremely and perhaps prohibitively time-intensive. 
This necessitates the development of more efficient techniques to model strong lensing systems. 
There have been some efforts to develop automated lens modeling techniques \citep[e.g.][]{Nightingale2016}, though this is a nascent field with many approaches that have not yet been explored. 

In this paper, we leverage Simulation-Based Inference (see \citep[]{CB2019} for an overview) to model challenging lensing systems that we expect to discover in current and future ground-based surveys. 
At the time of this work, related studies have also begun exploring the efficacy of SBI for strong lens modeling \citep[]{Legin2021,levasseur2017uncertainties, pearson2021strong, wagner2021hierarchical, park2021large}. These efforts have thus far been focused on space-based imaging data. Since we expect new lens discoveries to come from both ground- and space-based survey data, we present a novel and complementary application of both NPE and BNN methods on simulated ground-based imaging data, which typically have a larger noise profile than space-based imaging due to the effects of sky background noise and atmospheric blurring. For example. space-based HST F814W $i$-band images have sky-brightness magnitudes of order $\sim$22 mag/arcsec$^2$ and point spread function (PSF) full-width half maximum (FWHM) of order $\sim$0''.1 and pixel scales of order $\sim$0''.05, whereas ground-based imaging from the equivalent $i$-band filter on DES has typical sky-brightness of $\sim$20 mag/arcsec$^2$, PSF FWHMs of order $\sim$1'' and pixel-scale of 0.263''. The effects of these differing noise characteristics on the performance of deep learning models is not yet well-understood and is grounds for further exploration.
 \vspace{-2mm}
\section{Simulations of Strong Lenses}
\vspace{-2mm}
The population of strong lenses predicted from future surveys is dominated by galaxy-galaxy lenses, where a single source is lensed by a single gravitational object. 
Of these lenses, the vast majority are predicted to be early-type (elliptical) galaxies, as they comprise approximately 80$\%$ of the total lensing probability. 
Elliptical galaxy lenses are well-approximated by Singular Isothermal Ellipsoid (SIE) mass profiles, and 
accurate modeling of these lenses is essential for further science. 
We use \deeplenstronomy\footnote{\url{https://github.com/deepskies/deeplenstronomy}}~\cite{MN2021}, which is built on \lenstronomy\footnote{\url{https://github.com/sibirrer/lenstronomy}}, to simulate strong lensing systems for training and testing~\cite{BA2018,BS2021}.
These packages together provide state-of-the-art simulations of lensing systems, including high-fidelity ray-tracing of light from source objects around lenses, emulation of cosmic survey observations, and the generation of sample distributions. 
We use \deeplenstronomy's built-in emulation of DES-like survey conditions, which uses the DES's image pixel scale, wideband filters, observational noise, and PSF-blurring. 
We simulate DES-like observing conditions by drawing on empirical distributions of sky brightness and seeing from Figures 4 and 5 of \cite{DESdatarelease1}.

We generate 800,000 galaxy-galaxy strong lensing systems to train both the NPE and BNN algorithms. Our model includes $12$ parameters that describe both the Sersic source light, SIE lens mass and external shear in the galaxy-galaxy lens system. We draw from a uniform prior distribution to generate the training set.
A test set of 1000 simulated strong lensing systems was used to quantify the performance of the two methods. 
We restrict the test-set prior range (see Table~\ref{table:ensemblestats} in the Appendix for both training and test set prior ranges) to ensure that the performance of the algorithms near the limits of the test-set range is not degraded due to a lack of examples in the training set for the algorithms to learn. 
Figure~\ref{fig:lenses} in the Appendix shows $20$ randomly selected lenses from the test set. All the images used for training and testing are single-band images that are 32x32 pixels in size. 
\vspace{-2mm}
\section{Methodology}
\label{sec:sgl}
\vspace{-2mm}
\textbf{Neural Posterior Estimation (NPE):} In statistical inference, the posterior is given as $p(\theta | X) = (p(X | \theta) * p(\theta))/p(X)$,
where $p(X | \theta)$ denotes the likelihood function, $p(\theta)$ the prior and $p(X)$ the marginal likelihood.
Due to the intractability of the likelihood function, in SBI, the likelihood function is replaced with simulation-based outputs. 
Here, we use the NPE method described in \citep{greenberg2019automatic} and implemented in  \sbi\footnote{\url{https://www.mackelab.org/sbi/}} \citep{tejero-cantero2020sbi} to train a neural density estimator and directly infer the lens parameter posteriors from simulated single-band lens images. 
NPE uses three inputs -- a mechanistic model (simulator), priors on parameters of the model, and data (or corresponding summary statistics). 
The priors are utilized to sample the parameters and simulate synthetic data that is passed to a neural network (density estimator).
The network is trained to learn the relation between simulated data and the underlying parameters and later deployed on the empirical data to obtain the posterior distribution. 
NPE avoids likelihood calculations and uses simulations to train the network and get the relevant parameters.
We use Masked Autoregressive Flow~\citep[MAF;][]{papamakarios2017masked} as the density estimator and a custom Convolutional Neural Network as the embedding network that learns the summaries of the high-dimensional output of the simulations. 
 MAF uses normalizing flows~\citep{rezende2015variational} to convert a base distribution into a complex target distribution via a set of invertible transforms and a tractable Jacobian, followed by autoregressive density estimation.
In autoregressive density estimation~\citep{uria2016neural}, the joint density $p(\theta | \hat{x})$ is decomposed into a sequential product of conditional densities, where the decomposition is given as  $p(\theta | \hat{x}) = \prod_k p(\theta_k |\theta_{1:k-1}, \hat{x})$.
The MAF is constructed using stacked autoregressive transformations with a different ordering of $\theta$ in each transformation. Here, we use MAF with 400 hidden units and 20 transformations each.
The embedding network has six convolutional layers, one dense (fully connected) layer and the output layer (see Table \ref{table:embedding_net} for embedding network architecture details). 
The density estimator and embedding network are trained together for the inference step.

\textbf{Bayesian Neural Network (BNN):} With BNNs~\cite{NR1996}, the deterministic weights of the model are replaced by probability distributions, which can be used subsequently to provide a measure of how (un)certain a model is in its predictions. 
The objective of training a BNN is to find the posterior distribution $p(w|X,Y)$ of the weights $w$, given the training datasets ${\cal D} =(X,Y)$, where $X$ are the inputs and $Y$ are the corresponding labels. 
Inferring the posterior with BNNs is a difficult task as it includes an integral over all model weights. Here we use a common approach called \textit{variational inference} to model the posterior using a simple variational distribution $q(w|\theta)$ such as a Gaussian, and trying to fit the distribution’s parameters $\theta$ to be as close as possible to the true posterior $p(w|{\cal D})$~\cite{GR2011,SL2019}. This is done by minimising the Kullback-Leibler (KL) divergence~\cite{KL1951} between the true and variational posterior probability distributions. BNNs are capable of capturing both \textit{aleatoric} (statistical - related to the intrinsic randomness of the data-generating or measurement process) and \textit{epistemic} (systematic - related to the model; reducible with more data or better model) uncertainties~\cite{DH2017,KG2017}. Here, we choose to compare the NPE results with a BNN trained on the same training dataset of 800,000 images. We train our BNN ($8$ convolutional layers and $4$ dense layers, see Appendix, Table~\ref{table:BNN} for the full BNN architecture) for $550$ epochs, using the Evidence Lower Bound (ELBO) loss that is comprised of the negative log-likelihood loss and the KL divergence. We use the Adadelta optimizer~\cite{Z2012} with an initial learning rate of 0.1.


\section{Results}
\begin{figure}[ht]
\begin{center}
\includegraphics[width=0.49\linewidth]
{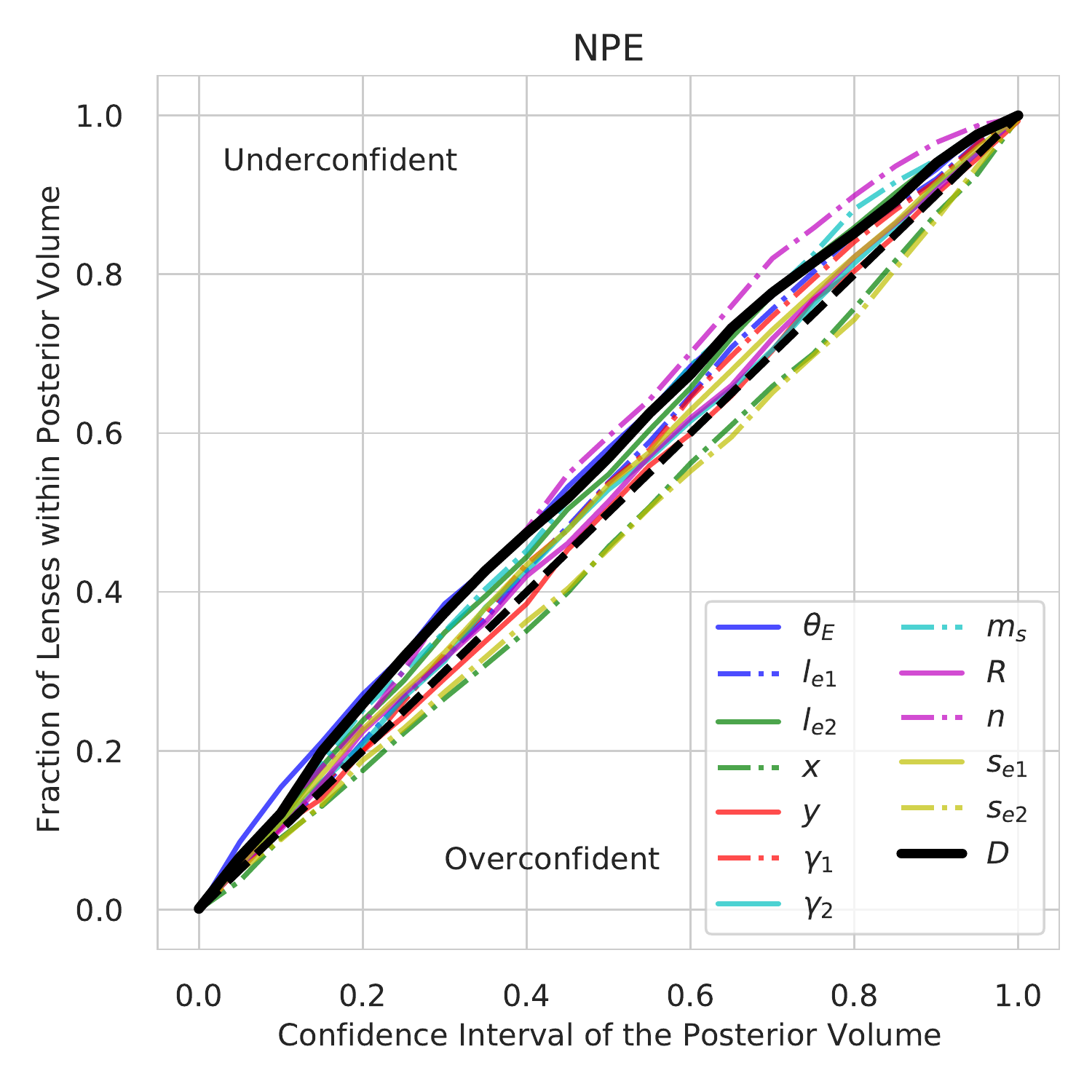}
\includegraphics[width=0.49\linewidth]
{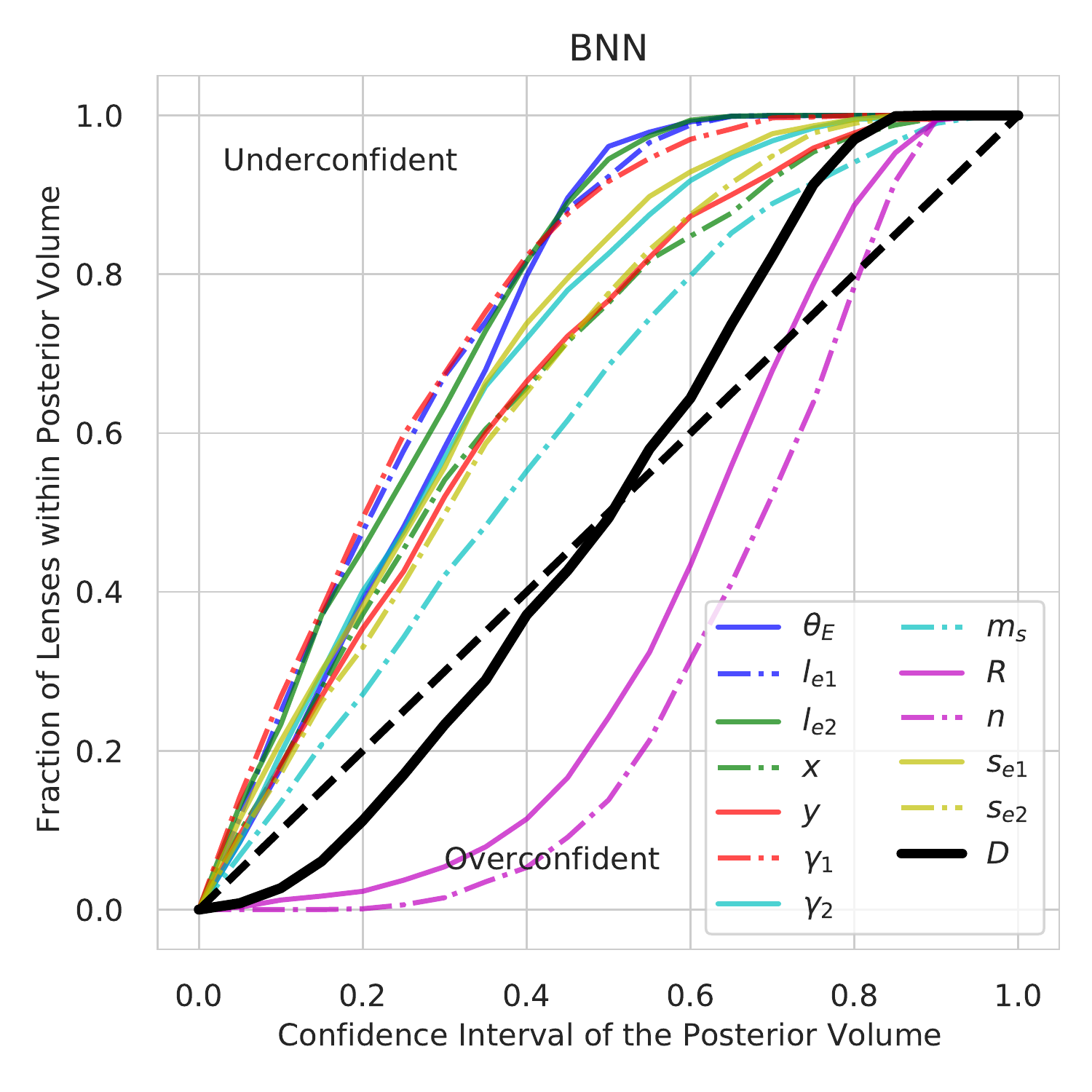}
\vspace{-4mm}
\caption{Posterior coverage plots of each of the 12 lens parameters inferred using the NPE (left) and BNN (right) methods for the same test set. The 1:1 line indicates perfect uncertainty calibration and is indicated by a dashed black line. The solid black line represents the distance metric described in Eq. 18 of \citep{wagner2021hierarchical}, which combines all lens parameters into a single distance metric while accounting for the empirical covariance between parameters. The NPE outperforms the BNN in every individual parameter and in the combined metric shown here.}
\label{fig:poscoverage}
\end{center}
\end{figure}

We characterize the performance of our trained NPE and BNN models on the same ensemble of 1000 simulated lens images in our test set. Both NPE and BNN models were trained using one Nvidia RTX A5000 GPU and training took approximately 6 hours in each case. We evaluate the performance of each model by quantifying (1) how well-calibrated each method is at inferring the correct posterior of the inferred parameter values, (2) how accurate the inferred best-fit posterior values of both methods are compared to the true values, and (3) hot the models perform on an out-of-distribution test set. The code and dataset used in this paper can be found in our github repository\footnote{\url{https://github.com/deepskies/DeepSLEEP}}.

\textbf{Uncertainty Calibration:} For a posterior distribution to be well-calibrated, it must contain the true value x\% of the time in x\% of the posterior probability volume. For example, a posterior distribution with 68\% (1$\sigma$ equivalent) confidence intervals should contain the true value within those confidence intervals 68\% of the time. To evaluate how well-calibrated the inferred posteriors of both methods are, we calculate the posterior coverage of both BNN and NPE methods. For the test set of 1000 images, we calculate for a given confidence interval of the posterior volume, what fraction of the true values fall within the interval. The results are shown in Figure~\ref{fig:poscoverage}.

\begin{figure}[ht]
\begin{center}
\centerline{\includegraphics[width=0.9\linewidth]
{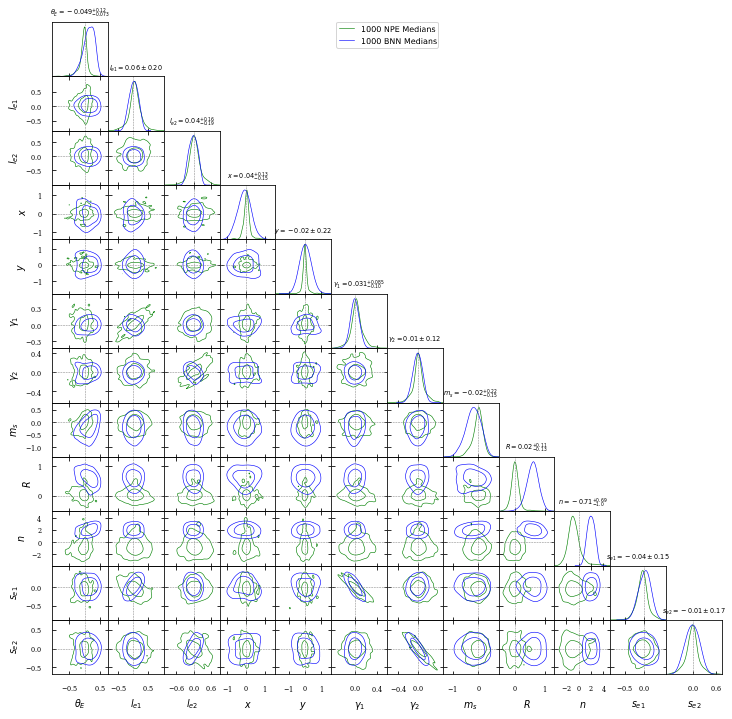}}
\vspace{-5mm}
\caption{Scatter plot matrix of the differences between the best-fit posterior values and true value for 1000 test images for NPE (green) and BNN (blue) methods with lens light subtracted. The contours show approximate 68th and 95th percentile uncertainties in the scatter. The dotted lines indicate the (0,0) point. Both methods produce a similar contour area suggesting roughly similar model precision, but BNN exhibits systematic bias for the source Sersic $R$ and $n$ parameters.}
\label{fig:1000uncertainties}
\end{center}
\end{figure}


\textbf{Ensemble Statistics of the Best-fit Values:} To characterize the statistical performance of the NPE and BNN methods over the full test set, we use the scatter in the difference between the best-fit posterior values from the true values of each lens parameter as a summary statistic. For each of the 1000 lenses in the test set, we use NPE and BNN methods to infer the lens parameter posteriors. We then subtract the best-fit values from the true value of the lens parameters for each lens. Figure~\ref{fig:1000uncertainties} shows the resulting scatter plot matrix for both NPE and BNN methods. The scatter characterizes the average performance of the method across the entire test set. If the method is completely accurate (i.e. it exactly predicts the true value for every single lens), we expect a single point at (0,0). If the method is systematically biased, we expect the scatter to shift away from the origin and if the method is not precise, we expect a large scatter in the distribution. Both methods produce a similar contour area, which suggests that both methods are roughly similarly precise. However, the BNN method appears off-center from the (0,0) point in a number of parameters, most notably the $R$ and $n$ parameters. This indicates a systematic bias in the BNN's inference of those lens parameter values. The values of the contours for each lens parameter are shown in Table~\ref{table:ensemblestats} in the Appendix.

\textbf{Out of Distribution Performance:}
To test the robustness of our SBI models, we apply it to an out-of-distribution (OOD) test set of images that were generated using different parameter distributions than that used for the training set. This is to emulate real world usage, as the distribution of hyperparameters in populations of real lens images are likely to differ from that used in training.
To evaluate the performances of our models on test sets with parameter distributions that differ from that of the training set, we generate a test set of 1000 images, but with narrower and shifted Gaussian priors, described in Table \ref{table:ensemblestats}. The results are shown in Figures \ref{fig:poscoverageood} and \ref{fig:1000uncertaintiesood} in the appendix. We find that having a test set generated with different distributions of priors does not significantly affect our conclusions for the NPE model. However, for the BNN, the best-fit values are more biased.  
\vspace{-2mm}
\section{Discussion}
\vspace{-2mm}
We successfully demonstrate the application of two SBI methods (NPE and BNN) to the problem of automated gravitational lens inference from simulated ground-based imaging data. We note that given the same training and test sets and similar computational costs, the NPE model significantly outperforms the BNN, both in the calibration of the uncertainties as well as the accuracy of the best-fit parameter values of the lens model. The BNN model's performance may be improved further by increasing the size of the training set or increasing the architecture complexity. However, both options will significantly increase computational costs. Nevertheless, the amortization of computational costs makes both options more economical than conventional MCMC techniques. Assuming an optimistic estimate of 10 minutes of MCMC computation time per lens, an analysis of the same test set of 1000 lenses would require around 166 hours of computation time, compared to the 6 hours it took to train both SBI models. This disparity in computational costs will only increase with larger test set sizes as SBI inferences are fast once the models have been trained.

\section{Broader Impact}
In the fields of astrophysics and cosmology, as well as other domains where uncertainty quantification is important, our work will enable automated parameter inference in cases where obtaining appropriate and credible uncertainty estimates of such parameter values is paramount.
While the scope of the work presented here is fairly narrow and applied specifically to images of strong lensing systems, it is broadly related to the much more general topic of generative modeling, uncertainty quantification and bayesian inference, all of which has already had a significant impact on society - both good and bad. We recognize that the broad use of these techniques without consideration of how complex real world data is and how it's potential impact can lead to unintended consequences and considerable harm. We encourage caution when trying to apply these techniques in such situations.

\begin{ack}



This manuscript has been supported by Fermi Research Alliance, LLC under Contract No. DE-AC02-07CH11359 with the U.S.\ Department of Energy (DOE), Office of Science, Office of High Energy Physics and by Subcontract 6749003 at the University of Chicago. 
The authors of this paper have committed themselves to performing this work in an equitable, inclusive, and just environment, and we hold ourselves accountable, believing that the best science is contingent on a good research environment.

We acknowledge the Deep Skies Lab as a community of multi-domain experts and collaborators who have facilitated an environment of open discussion, idea-generation, and collaboration. This community was important for the development of this project.

Furthermore, we also thank the anonymous referees whose comments helped improve this manuscript.

\textbf{Author Contributions:} 
J.~Poh: \textit{Conceptualization, Methodology, Software, Validation, Formal Analysis, Investigation, Data Curation, Visualization, Writing of original draft, Writing - Review \& Editing};
A.~Samudre: \textit{Formal Analysis, Investigation, Simulations, Methodology, Software, Data Curation, Validation, Visualization, Writing of original draft, Writing - Review \& Editing}
A.~\'Ciprijanovi\'c: \textit{Formal analysis, Investigation, Methodology, Project administration, Software, Supervision, Writing of original draft, Writing - Review \& Editing}; 
B.~Nord: \textit{Conceptualization, Methodology, Investigation, Supervision, Resources, Writing - Review \& Editing}; 
G.~Khullar: \textit{Formal Analysis, Software}
D.~Tanoglidis: \textit{Software}
J.~A.~Frieman: \textit{Supervision, Resources.}
\end{ack}

\medskip
\small

\bibliographystyle{plainnat}
\bibliography{main}

\begin{thebibliography}{35}
\providecommand{\natexlab}[1]{#1}
\providecommand{\url}[1]{\texttt{#1}}
\expandafter\ifx\csname urlstyle\endcsname\relax
  \providecommand{\doi}[1]{doi: #1}\else
  \providecommand{\doi}{doi: \begingroup \urlstyle{rm}\Url}\fi

\bibitem[{Abbott} et~al.(2018){Abbott}, {Abdalla}, {Allam}, {Amara}, {Annis},
  {Asorey}, {Avila}, {Ballester}, et~al., and {NOAO Data Lab}]{DESdatarelease1}
T.~M.~C. {Abbott}, F.~B. {Abdalla}, S.~{Allam}, A.~{Amara}, J.~{Annis},
  J.~{Asorey}, S.~{Avila}, O.~{Ballester}, et~al., and {NOAO Data Lab}.
\newblock {The Dark Energy Survey: Data Release 1}.
\newblock \emph{\apjs}, 239\penalty0 (2):\penalty0 18, December 2018.
\newblock \doi{10.3847/1538-4365/aae9f0}.

\bibitem[{Aihara} et~al.(2018){Aihara}, {Arimoto}, {Armstrong}, {Arnouts},
  {Bahcall}, {Bickerton}, {Bosch}, {Bundy}, and et~al.]{HSC2018}
Hiroaki {Aihara}, Nobuo {Arimoto}, Robert {Armstrong}, St{\'e}phane {Arnouts},
  Neta~A. {Bahcall}, Steven {Bickerton}, James {Bosch}, Kevin {Bundy}, and
  et~al.
\newblock {The Hyper Suprime-Cam SSP Survey: Overview and survey design}.
\newblock \emph{\pasj}, 70:\penalty0 S4, January 2018.
\newblock \doi{10.1093/pasj/psx066}.

\bibitem[Auger et~al.(2010)Auger, Treu, Bolton, Gavazzi, Koopmans, Marshall,
  Moustakas, and Burles]{Auger2010a}
M.~W. Auger, T.~Treu, A.~S. Bolton, R.~Gavazzi, L.~V.~E. Koopmans, P.~J.
  Marshall, L.~A. Moustakas, and S.~Burles.
\newblock {The Sloan Lens ACS Survey. X. Stellar, Dynamical, and Total Mass
  Correlations of Massive Early-type Galaxies}.
\newblock \emph{The Astrophysical Journal, Volume 724, Issue 1, pp. 511-525
  (2010).}, 724:\penalty0 511--525, jul 2010.
\newblock ISSN 0004-637X.
\newblock \doi{10.1088/0004-637X/724/1/511}.

\bibitem[Barnabe et~al.(2011)Barnabe, Czoske, Koopmans, Treu, and
  Bolton]{Barnabe2011}
Matteo Barnabe, Oliver Czoske, Leon V.~E. Koopmans, Tommaso Treu, and Adam~S.
  Bolton.
\newblock {Two-dimensional kinematics of SLACS lenses: III. Mass structure and
  dynamics of early-type lens galaxies beyond z {\~{}} 0.1}.
\newblock \emph{Monthly Notices of the Royal Astronomical Society, Volume 415,
  Issue 3, pp. 2215-2232.}, 415:\penalty0 2215--2232, feb 2011.
\newblock ISSN 0035-8711.
\newblock \doi{10.1111/j.1365-2966.2011.18842.x}.

\bibitem[{Birrer} and {Amara}(2018)]{BA2018}
Simon {Birrer} and Adam {Amara}.
\newblock {lenstronomy: Multi-purpose gravitational lens modelling software
  package}.
\newblock \emph{Physics of the Dark Universe}, 22:\penalty0 189--201, December
  2018.
\newblock \doi{10.1016/j.dark.2018.11.002}.

\bibitem[{Birrer} et~al.(2021){Birrer}, {Shajib}, {Gilman}, {Galan}, {Aalbers},
  {Millon}, {Morgan}, {Pagano}, {Park}, {Teodori}, {Tessore}, {Ueland}, {Van de
  Vyvere}, {Wagner-Carena}, {Wempe}, {Yang}, {Ding}, {Schmidt}, {Sluse},
  {Zhang}, and {Amara}]{BS2021}
Simon {Birrer}, Anowar {Shajib}, Daniel {Gilman}, Aymeric {Galan}, Jelle
  {Aalbers}, Martin {Millon}, Robert {Morgan}, Giulia {Pagano}, Ji~{Park}, Luca
  {Teodori}, Nicolas {Tessore}, Madison {Ueland}, Lyne {Van de Vyvere},
  Sebastian {Wagner-Carena}, Ewoud {Wempe}, Lilan {Yang}, Xuheng {Ding}, Thomas
  {Schmidt}, Dominique {Sluse}, Ming {Zhang}, and Adam {Amara}.
\newblock {lenstronomy II: A gravitational lensing software ecosystem}.
\newblock \emph{The Journal of Open Source Software}, 6\penalty0 (62):\penalty0
  3283, June 2021.
\newblock \doi{10.21105/joss.03283}.

\bibitem[{Cranmer} et~al.(2019){Cranmer}, {Brehmer}, and {Louppe}]{CB2019}
Kyle {Cranmer}, Johann {Brehmer}, and Gilles {Louppe}.
\newblock {The frontier of simulation-based inference}.
\newblock \emph{arXiv e-prints}, art. arXiv:1911.01429, November 2019.

\bibitem[{Dark Energy Survey Collaboration} et~al.(2016){Dark Energy Survey
  Collaboration}, {Abbott}, {Abdalla}, {Aleksi{\'c}}, {Allam}, {Amara},
  {Bacon}, {Balbinot}, {Banerji}, {Bechtol}, {Benoit-L{\'e}vy}, and
  et~al.]{DES2016}
{Dark Energy Survey Collaboration}, T.~{Abbott}, F.~B. {Abdalla},
  J.~{Aleksi{\'c}}, S.~{Allam}, A.~{Amara}, D.~{Bacon}, E.~{Balbinot},
  M.~{Banerji}, K.~{Bechtol}, A.~{Benoit-L{\'e}vy}, and et~al.
\newblock {The Dark Energy Survey: more than dark energy - an overview}.
\newblock \emph{\mnras}, 460\penalty0 (2):\penalty0 1270--1299, August 2016.
\newblock \doi{10.1093/mnras/stw641}.

\bibitem[{Depeweg} et~al.(2017){Depeweg}, {Hern{\'a}ndez-Lobato},
  {Doshi-Velez}, and {Udluft}]{DH2017}
Stefan {Depeweg}, Jos{\'e}~Miguel {Hern{\'a}ndez-Lobato}, Finale {Doshi-Velez},
  and Steffen {Udluft}.
\newblock {Decomposition of Uncertainty in Bayesian Deep Learning for Efficient
  and Risk-sensitive Learning}.
\newblock \emph{arXiv e-prints}, art. arXiv:1710.07283, October 2017.

\bibitem[{Graves}(2011)]{GR2011}
Alex {Graves}.
\newblock Practical variational inference for neural networks.
\newblock In J.~Shawe-Taylor, R.~Zemel, P.~Bartlett, F.~Pereira, and K.Q.
  Weinberger, editors, \emph{Advances in Neural Information Processing
  Systems}, volume~24. Curran Associates, Inc., 2011.
\newblock URL
  \url{https://proceedings.neurips.cc/paper/2011/file/7eb3c8be3d411e8ebfab08eba5f49632-Paper.pdf}.

\bibitem[Greenberg et~al.(2019)Greenberg, Nonnenmacher, and
  Macke]{greenberg2019automatic}
David Greenberg, Marcel Nonnenmacher, and Jakob Macke.
\newblock Automatic posterior transformation for likelihood-free inference.
\newblock In \emph{International Conference on Machine Learning}, pages
  2404--2414. PMLR, 2019.

\bibitem[{Ivezi{\'c}} et~al.(2019){Ivezi{\'c}}, {Kahn}, {Tyson}, {Abel},
  {Acosta}, {Allsman}, {Alonso}, {AlSayyad}, and et~al.]{IK2019}
{\v{Z}}eljko {Ivezi{\'c}}, Steven~M. {Kahn}, J.~Anthony {Tyson}, Bob {Abel},
  Emily {Acosta}, Robyn {Allsman}, David {Alonso}, Yusra {AlSayyad}, and et~al.
\newblock {LSST}: From science drivers to reference design and anticipated data
  products.
\newblock \emph{ApJ}, 873\penalty0 (2):\penalty0 111, March 2019.
\newblock \doi{10.3847/1538-4357/ab042c}.

\bibitem[{Kendall} and {Gal}(2017)]{KG2017}
A.~{Kendall} and Y~{Gal}.
\newblock {What Uncertainties Do We Need in Bayesian Deep Learning for Computer
  Vision?}
\newblock In \emph{arXiv:1703.04977}, March 2017.

\bibitem[Kullback and Leibler(1951)]{KL1951}
Solomon Kullback and Richard~A Leibler.
\newblock On information and sufficiency.
\newblock \emph{The annals of mathematical statistics}, 22\penalty0
  (1):\penalty0 79--86, 1951.

\bibitem[{Legin} et~al.(2021){Legin}, {Hezaveh}, {Perreault Levasseur}, and
  {Wandelt}]{Legin2021}
Ronan {Legin}, Yashar {Hezaveh}, Laurence {Perreault Levasseur}, and Benjamin
  {Wandelt}.
\newblock {Simulation-Based Inference of Strong Gravitational Lensing
  Parameters}.
\newblock \emph{arXiv e-prints}, art. arXiv:2112.05278, December 2021.

\bibitem[Levasseur et~al.(2017)Levasseur, Hezaveh, and
  Wechsler]{levasseur2017uncertainties}
Laurence~Perreault Levasseur, Yashar~D Hezaveh, and Risa~H Wechsler.
\newblock Uncertainties in parameters estimated with neural networks:
  Application to strong gravitational lensing.
\newblock \emph{The Astrophysical Journal Letters}, 850\penalty0 (1):\penalty0
  L7, 2017.

\bibitem[{Morgan} et~al.(2021){Morgan}, {Nord}, {Birrer}, {Lin}, and
  {Poh}]{MN2021}
Robert {Morgan}, Brian {Nord}, Simon {Birrer}, Joshua {Lin}, and Jason {Poh}.
\newblock {deeplenstronomy: A dataset simulation package for strong
  gravitational lensing}.
\newblock \emph{The Journal of Open Source Software}, 6\penalty0 (58):\penalty0
  2854, February 2021.
\newblock \doi{10.21105/joss.02854}.

\bibitem[{Neal}(1996)]{NR1996}
R.~M {Neal}.
\newblock \emph{Bayesian Learning for Neural Networks}, volume 118.
\newblock Springer Science \& Business Media, 1996.
\newblock ISBN 978-0-387-94724-2.

\bibitem[Newman et~al.(2012{\natexlab{a}})Newman, Treu, Ellis, and
  Sand]{Newman2012}
Andrew~B. Newman, Tommaso Treu, Richard~S. Ellis, and David~J. Sand.
\newblock {The Density Profiles of Massive, Relaxed Galaxy Clusters. II.
  Separating Luminous and Dark Matter in Cluster Cores}.
\newblock \emph{The Astrophysical Journal, Volume 765, Issue 1, article id. 25,
  12 pp. (2013).}, 765, sep 2012{\natexlab{a}}.
\newblock ISSN 0004-637X.
\newblock \doi{10.1088/0004-637X/765/1/25}.

\bibitem[Newman et~al.(2012{\natexlab{b}})Newman, Treu, Ellis, Sand, Nipoti,
  Richard, and Jullo]{Newman2012a}
Andrew~B. Newman, Tommaso Treu, Richard~S. Ellis, David~J. Sand, Carlo Nipoti,
  Johan Richard, and Eric Jullo.
\newblock {The Density Profiles of Massive, Relaxed Galaxy Clusters. I. The
  Total Density Over Three Decades in Radius}.
\newblock \emph{The Astrophysical Journal, Volume 765, Issue 1, article id. 24,
  35 pp. (2013).}, 765, sep 2012{\natexlab{b}}.
\newblock ISSN 0004-637X.
\newblock \doi{10.1088/0004-637X/765/1/24}.

\bibitem[Newman et~al.(2015)Newman, Ellis, and Treu]{Newman2015}
Andrew~B. Newman, Richard~S. Ellis, and Tommaso Treu.
\newblock {Luminous and Dark Matter Profiles from Galaxies to Clusters:
  Bridging the Gap with Group-scale Lenses}.
\newblock \emph{The Astrophysical Journal, Volume 814, Issue 1, article id. 26,
  28 pp. (2015).}, 814, 2015.
\newblock ISSN 0004-637X.
\newblock \doi{10.1088/0004-637X/814/1/26}.

\bibitem[Nightingale(2016)]{Nightingale2016}
James~J.N. Nightingale.
\newblock \emph{{AutoLens: automated modeling of a strong lens's light, mass
  and source}}.
\newblock PhD thesis, University of Nottingham, dec 2016.
\newblock URL \url{http://eprints.nottingham.ac.uk/38507/}.

\bibitem[Papamakarios et~al.(2017)Papamakarios, Pavlakou, and
  Murray]{papamakarios2017masked}
George Papamakarios, Theo Pavlakou, and Iain Murray.
\newblock Masked autoregressive flow for density estimation.
\newblock \emph{arXiv preprint arXiv:1705.07057}, 2017.

\bibitem[Park et~al.(2021)Park, Wagner-Carena, Birrer, Marshall, Lin, Roodman,
  Collaboration, et~al.]{park2021large}
Ji~Won Park, Sebastian Wagner-Carena, Simon Birrer, Philip~J Marshall, Joshua
  Yao-Yu Lin, Aaron Roodman, LSST Dark Energy~Science Collaboration, et~al.
\newblock Large-scale gravitational lens modeling with bayesian neural networks
  for accurate and precise inference of the hubble constant.
\newblock \emph{The Astrophysical Journal}, 910\penalty0 (1):\penalty0 39,
  2021.

\bibitem[Pearson et~al.(2021)Pearson, Maresca, Li, and Dye]{pearson2021strong}
James Pearson, Jacob Maresca, Nan Li, and Simon Dye.
\newblock Strong lens modelling: comparing and combining bayesian neural
  networks and parametric profile fitting.
\newblock \emph{Monthly Notices of the Royal Astronomical Society},
  505\penalty0 (3):\penalty0 4362--4382, 2021.

\bibitem[Rezende and Mohamed(2015)]{rezende2015variational}
Danilo Rezende and Shakir Mohamed.
\newblock Variational inference with normalizing flows.
\newblock In \emph{International conference on machine learning}, pages
  1530--1538. PMLR, 2015.

\bibitem[{Shajib} et~al.(2020){Shajib}, {Birrer}, {Treu}, {Agnello},
  {Buckley-Geer}, {Chan}, {Christensen}, {Lemon}, and et~al.]{Shajib2020}
A.~J. {Shajib}, S.~{Birrer}, T.~{Treu}, A.~{Agnello}, E.~J. {Buckley-Geer},
  J.~H.~H. {Chan}, L.~{Christensen}, C.~{Lemon}, and et~al.
\newblock {STRIDES: a 3.9 per cent measurement of the Hubble constant from the
  strong lens system DES J0408-5354}.
\newblock \emph{\mnras}, 494\penalty0 (4):\penalty0 6072--6102, June 2020.
\newblock \doi{10.1093/mnras/staa828}.

\bibitem[Shridhar et~al.(2019)Shridhar, Laumann, and Liwicki]{SL2019}
Kumar Shridhar, Felix Laumann, and Marcus Liwicki.
\newblock A comprehensive guide to bayesian convolutional neural network with
  variational inference.
\newblock \emph{arXiv preprint arXiv:1901.02731}, 2019.

\bibitem[Suyu et~al.(2010)Suyu, Marshall, Auger, Hilbert, Blandford, Koopmans,
  Fassnacht, and Treu]{Suyu2010}
S.~H. Suyu, P.~J. Marshall, M.~W. Auger, S.~Hilbert, R.~D. Blandford, L.~V.~E.
  Koopmans, C.~D. Fassnacht, and T.~Treu.
\newblock {Dissecting the Gravitational lens B1608+656. II. Precision
  Measurements of the Hubble Constant, Spatial Curvature, and the Dark Energy
  Equation of State}.
\newblock \emph{The Astrophysical Journal, Volume 711, Issue 1, pp. 201-221
  (2010).}, 711:\penalty0 201--221, 2010.
\newblock ISSN 0004-637X.
\newblock \doi{10.1088/0004-637X/711/1/201}.

\bibitem[Suyu et~al.(2013)Suyu, Auger, Hilbert, Marshall, Tewes, Treu,
  Fassnacht, Koopmans, Sluse, Blandford, Courbin, and Meylan]{Suyu2013}
S.~H. Suyu, M.~W. Auger, S.~Hilbert, P.~J. Marshall, M.~Tewes, T.~Treu, C.~D.
  Fassnacht, L.~V.~E. Koopmans, D.~Sluse, R.~D. Blandford, F.~Courbin, and
  G.~Meylan.
\newblock {Two Accurate Time-delay Distances from Strong Lensing: Implications
  for Cosmology}.
\newblock \emph{The Astrophysical Journal, Volume 766, Issue 2, article id. 70,
  19 pp. (2013).}, 766, 2013.
\newblock ISSN 0004-637X.
\newblock \doi{10.1088/0004-637X/766/2/70}.

\bibitem[Tejero-Cantero et~al.(2020)Tejero-Cantero, Boelts, Deistler,
  Lueckmann, Durkan, Gonçalves, Greenberg, and Macke]{tejero-cantero2020sbi}
Alvaro Tejero-Cantero, Jan Boelts, Michael Deistler, Jan-Matthis Lueckmann,
  Conor Durkan, Pedro~J. Gonçalves, David~S. Greenberg, and Jakob~H. Macke.
\newblock sbi: A toolkit for simulation-based inference.
\newblock \emph{Journal of Open Source Software}, 5\penalty0 (52):\penalty0
  2505, 2020.
\newblock \doi{10.21105/joss.02505}.
\newblock URL \url{https://doi.org/10.21105/joss.02505}.

\bibitem[Uria et~al.(2016)Uria, C{\^o}t{\'e}, Gregor, Murray, and
  Larochelle]{uria2016neural}
Benigno Uria, Marc-Alexandre C{\^o}t{\'e}, Karol Gregor, Iain Murray, and Hugo
  Larochelle.
\newblock Neural autoregressive distribution estimation.
\newblock \emph{The Journal of Machine Learning Research}, 17\penalty0
  (1):\penalty0 7184--7220, 2016.

\bibitem[Wagner-Carena et~al.(2021)Wagner-Carena, Park, Birrer, Marshall,
  Roodman, Wechsler, Collaboration, et~al.]{wagner2021hierarchical}
Sebastian Wagner-Carena, Ji~Won Park, Simon Birrer, Philip~J Marshall, Aaron
  Roodman, Risa~H Wechsler, LSST Dark Energy~Science Collaboration, et~al.
\newblock Hierarchical inference with bayesian neural networks: An application
  to strong gravitational lensing.
\newblock \emph{The Astrophysical Journal}, 909\penalty0 (2):\penalty0 187,
  2021.

\bibitem[Wen et~al.(2018)Wen, Vicol, Ba, Tran, and Grosse]{wen2018flipout}
Yeming Wen, Paul Vicol, Jimmy Ba, Dustin Tran, and Roger Grosse.
\newblock Flipout: Efficient pseudo-independent weight perturbations on
  mini-batches.
\newblock \emph{arXiv preprint arXiv:1803.04386}, 2018.

\bibitem[{Zeiler}(2012)]{Z2012}
Matthew~D. {Zeiler}.
\newblock {ADADELTA: An Adaptive Learning Rate Method}.
\newblock \emph{arXiv e-prints}, art. arXiv:1212.5701, December 2012.

\end{thebibliography}








\clearpage
\appendix

\section{Appendix}\label{sec:appendix}



\begin{figure}[ht]
\begin{center}
\centerline{\includegraphics[width=0.8\columnwidth]
{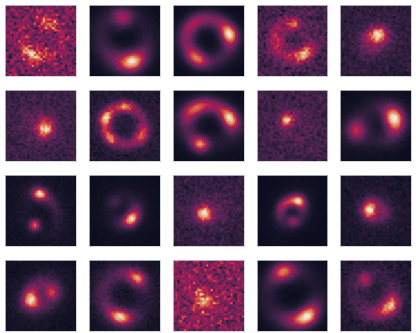}}
\caption{20 randomly selected lens-subtracted lens images from the test set.}
\label{fig:lenses}
\end{center}
\end{figure}

\begin{table}
  \centering
  \noindent\begin{minipage}[b]{0.99\columnwidth}
  \centering
    \caption{
    The architecture of the embedding network used in our SBI method. In the parameters column, `k' denotes the kernel size and `s' denotes the stride, in\_ft denotes the input features for the Linear (Dense) layer. We found in our experiments that having $4N$ summaries, where $N$ is the number of output parameters, results in good performance for our problem. Hence, for our 12-parameter output, we summarise the imaging data into $48$ summary parameters.
    }
  \label{table:embedding_net}
  \centering
  \begin{tabular}{lc c}
 \hline   Layer   &  Output shape   &  Parameters \\\hline \hline
  Conv2d    &  [-1, 8, 32, 32]  &  k=3, s=1\\ 
  \midrule
  BatchNorm2d & [-1, 8, 32, 32] & k=3, s=1\\ 
  \midrule
  Conv2d & [-1, 16, 32, 32] & k=3, s=1 \\
  \midrule
  BatchNorm2d & [-1, 16, 32, 32] & k=3, s=1\\ 
  \midrule
  MaxPool2d & [-1, 16, 16, 16] & k=2, s=2  \\
  \midrule
  Conv2d    &  [-1, 32, 16, 16]  &  k=3, s=1\\ 
  \midrule
  BatchNorm2d & [-1, 32, 16, 16] & k=3, s=1\\ 
  \midrule
  Conv2d & [-1, 32, 16, 16] & k=3, s=1\\
  \midrule
  BatchNorm2d & [-1, 32, 16, 16] & k=3, s=1\\ 
  \midrule
  MaxPool2d & [-1, 32, 8, 8] & k=2, s=2 \\
  \midrule
  Conv2d    &  [-1, 64, 8, 8]  &  k=3, s=1 \\ 
  \midrule
  BatchNorm2d & [-1, 64, 8, 8] & k=3, s=1\\ 
  \midrule
  Conv2d & [-1, 128, 8, 8] & k=3, s=1\\
  \midrule
  BatchNorm2d & [-1, 128, 8, 8] & k=3, s=1\\ 
  \midrule
  MaxPool2d & [-1, 128, 4, 4] & k=2, s=2 \\
  \midrule
  Linear & [-1, 48] & in\_ft=128*4*4\\
  \hline

\end{tabular}
\end{minipage}
\end{table}

\begin{table}
  \centering
  \noindent\begin{minipage}[b]{0.99\columnwidth}
  \centering
    \caption{
    The Bayesian Neural Network architecture. In the parameters column, `k' denotes the kernel size and `s' denotes the stride. Conv2dFlipout denotes the 2D convolution layer with Flipout estimator~\citep{wen2018flipout} and DenseFlipout denotes the Dense layer with Flipout estimator. MultivariateNormalTriL denotes the multivariate normal distribution
    }
  \label{table:BNN}
  \centering
  \begin{tabular}{lc c}
 \hline   Layer   &  Output shape   &  Parameters \\\hline \hline
  Conv2dFlipout &  [-1, 16, 32, 32]  &  k=3, s=1\\ 
  \midrule
  MaxPool2d & [-1, 16, 16, 16] & k=2, s=2\\ 
  \midrule
  Conv2dFlipout &  [-1, 32, 16, 16]  &  k=3, s=1 \\
  \midrule
  Conv2dFlipout &  [-1, 32, 16, 16]  &  k=3, s=1\\ 
  \midrule
  MaxPool2d & [-1, 32, 8, 8] & k=2, s=2 \\
  \midrule
  Conv2dFlipout &  [-1, 48, 8, 8]  &  k=3, s=1\\ 
  \midrule
  Conv2dFlipout &  [-1, 48, 8, 8]  &  k=3, s=1\\ 
  \midrule
  MaxPool2d & [-1, 48, 4, 4] & k=2, s=2\\ 
  \midrule
  Conv2dFlipout &  [-1, 64, 4, 4]  &  k=3, s=1\\ 
  \midrule
  Conv2dFlipout &  [-1, 64, 4, 4]  &  k=3, s=1\\ 
  \midrule
  Conv2dFlipout &  [-1, 64, 4, 4]  &  k=3, s=1\\ 
  \midrule
  MaxPool2d & [-1, 64, 2, 2] & k=2, s=2\\ 
  \midrule
  Flatten & [-1, 256] & - \\
  \midrule
  DenseFlipout & [-1, 2048] & - \\
  \midrule
  DenseFlipout & [-1, 512] & - \\
  \midrule
  DenseFlipout & [-1, 64] & - \\
  \midrule
  Dense & [-1, 90] & - \\
  \midrule
  MultivariateNormalTriL & [-1, 12] & - \\
  \hline

\end{tabular}
\end{minipage}
\end{table}


\begin{table}[t]
\noindent\begin{minipage}[b]{0.99\columnwidth}
  \caption{Average 68\% scatter in the difference between the best-fit lens parameters from the true values for the test set of 1000 lenses. The full list of model parameters are as follows: for the lens SIE mass profile, Einstein radius (\einsteinrad), ellipticity components ($le_{1}$,$le_{2}$), and lens-source offset ($x,y$). For the lens environment, components of the external shear ($\gamma_1$, $\gamma_2$). For the source Sersic light profile, apparent magnitude ($m_s$), half-light radius ($R$), Sersic index ($n$) and ellipticity ($se_1$,$se_2$). The test set and training set prior ranges for each parameter are provided in the rightmost two columns. All priors are uniform, except for those in the out-of-distribution (OOD) test set, which may also have normal ($N$) or log-normal ($N_{log}$) distributions.}   
  \label{table:ensemblestats}
  \centering
  \begin{tabular}{lcccccc}
 \hline   Parameter  &  NPE   &  BNN & Test Set Priors & Training Set Priors & OOD test\\ \toprule 
 \multicolumn{6}{c}{Lens Mass Parameters}\\
 \midrule
   \einsteinrad  ('')     &  $0.05 \pm 0.1$    &  $0.13 \pm 0.2$ & [0.5,3.0] & [0.3,4.0] & $N(1,0.2)$  \\ 
  $le_{2}$       & $0.06 \pm 0.2$       &   $0.02 \pm 0.2$  & [-0.2,0.2] & [-0.8,0.8] &$N(0.2,0.1)$\\ 
  $le_{2}$      &  $0.04 \pm 0.2$    &    $-0.01 \pm 0.2$ &[-0.2,0.2] & [-0.8,0.8] & $N(-0.2,0.1)$\\ 
   $x_\mathrm{}$ (")       &  $0.04\pm 0.2$    &   $0.09 \pm 0.4$  & [-1,1] & [-2,2] & $N(0.2,0.2)$\\ 
   $y_\mathrm{}$ (")        & $-0.02 \pm 0.2$      &  $0.03 \pm 0.4$ & [-1,1] & [-2,2] &$N(-0.2,0.2)$\\
 \midrule
 \multicolumn{6}{c}{Lens Environment Parameters}\\
 \midrule
 $\gamma_1$        &  $0.03 \pm 0.1$    &  $0.003 \pm 0.08$&  [-0.05,0.05] & [-0.8,0.8] & $N_{log}(-3,1)$ \\ 
 $\gamma_2$       & $ 0.01\pm 0.1$     &   $ -0.01 \pm 0.1$  & [-0.05,0.05] & [-0.8,0.8] & $N_{log}(-3,1)$ \\  \midrule
 \multicolumn{6}{c}{Source Light Parameters}\\
 \midrule
  $m_s$       & $ -0.02\pm 0.2 $     &    $ -0.2\pm 0.3$ & [19,24] & [18,25] & $N(22,1)$ \\ 
  $R$ (")       &  $ 0.02\pm 0.1$    &   $ 0.6\pm 0.2$ & [0.5,1.0] & [0.1,3.0] & $N(0.7,0.1)$ \\ 
   $n$        &  $ -0.7\pm 1.0$    &   $ 2.0\pm0.6 $ & [2,4] & [0.5,8.0] & $N(4,1)$\\ 
 $se_1$        &  $ -0.04\pm 0.2$    &   $ 0.03\pm 0.2$ & [-0.2,0.2] & [-0.8,0.8] &$N(-0.2,0.2)$\\ 
 $se_2$        & $ -0.01\pm 0.2$     & $ -0.01\pm 0.2$  & [-0.2,0.2] & [-0.8,0.8] &$N(0.2,0.2)$\\ \hline
\end{tabular}
\end{minipage}
\end{table}


\begin{figure}[ht]
\begin{center}
\includegraphics[width=0.49\linewidth]
{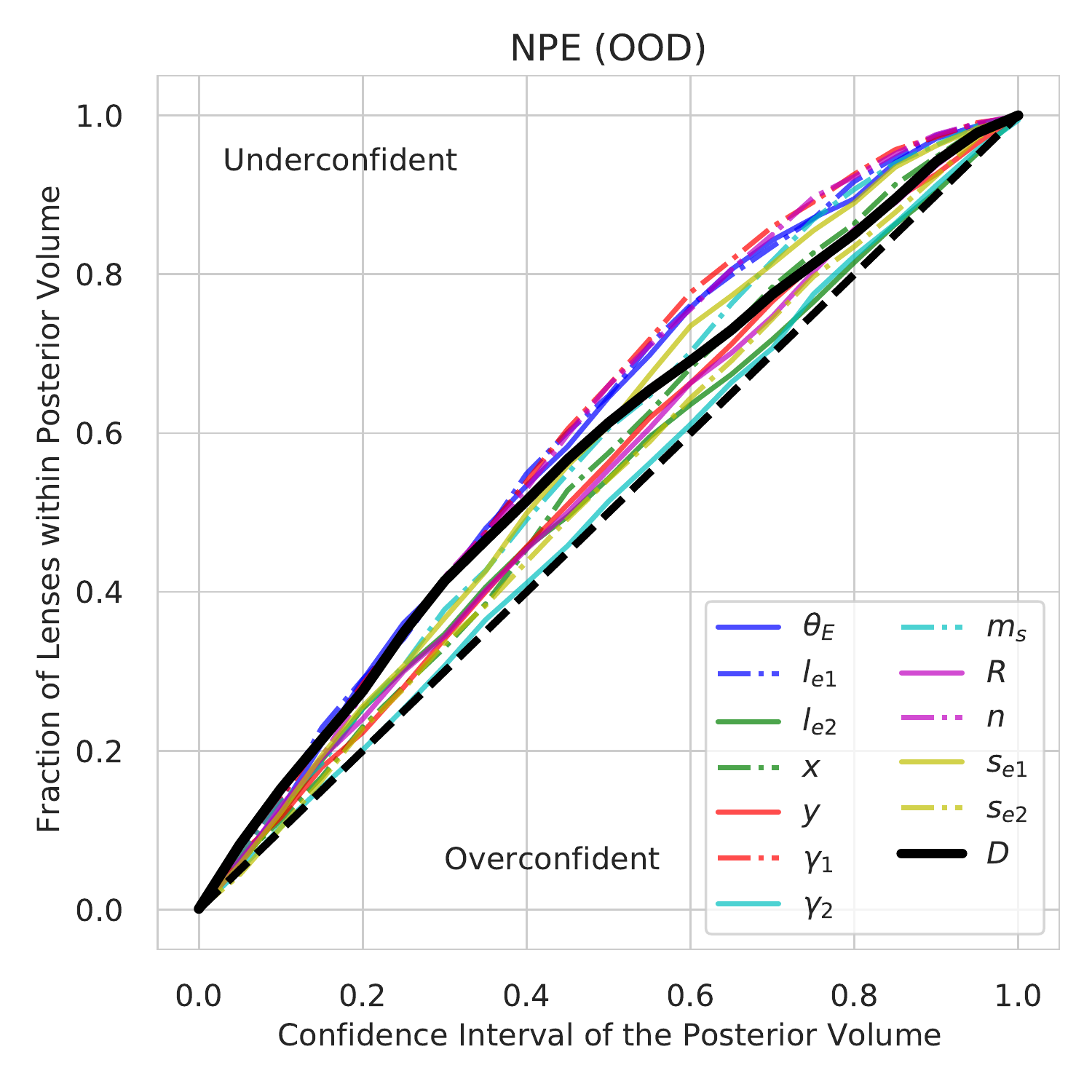}
\includegraphics[width=0.49\linewidth]
{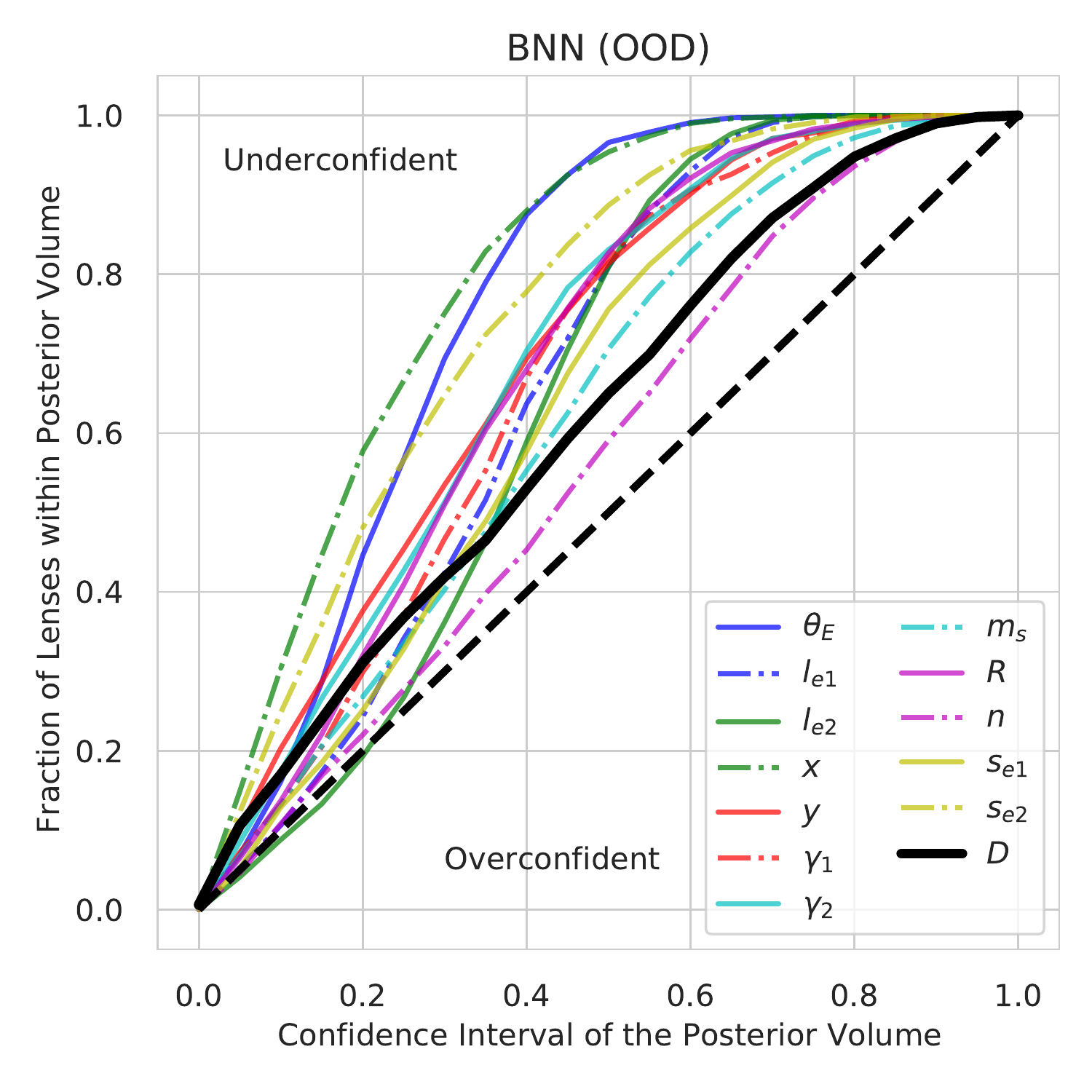}
\caption{Posterior coverage plots of each of the 12 lens parameters inferred using the NPE (left) and BNN (right) methods on an out-of-distribution test set with priors given in table \ref{table:ensemblestats}. The 1:1 line indicates perfect uncertainty calibration and is indicated by a dashed black line. The solid black line represents the distance metric described in Eq. 18 of \citep{wagner2021hierarchical}, which combines all lens parameters into a single distance metric while accounting for the empirical covariance between parameters.}
\label{fig:poscoverageood}
\end{center}
\end{figure}

\begin{figure}[ht]
\begin{center}
\centerline{\includegraphics[width=1.4\linewidth]
{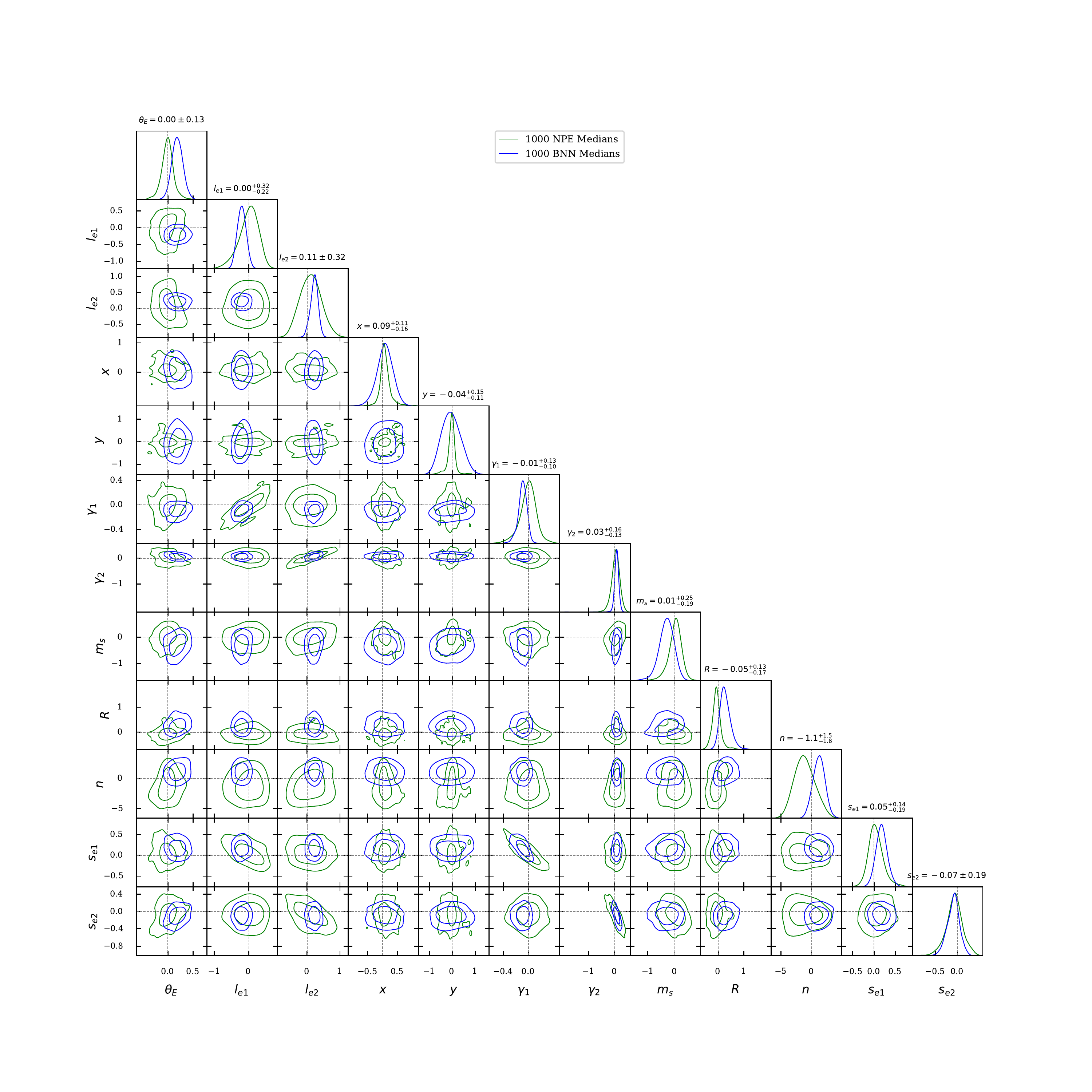}}
\caption{Scatter plot matrix of the differences between the best-fit posterior values and true value for 1000 test images for NPE (green) and BNN (blue) on an out-of-distribution test set with priors given in table \ref{table:ensemblestats}. The contours show approximate 68th and 95th percentile uncertainties in the scatter. The dotted lines indicate the (0,0) point. Both methods produce a similar contour area suggesting roughly similar model precision, but BNN exhibits systematic bias for the source Sersic $R$ and $n$ parameters.}
\label{fig:1000uncertaintiesood}
\end{center}
\end{figure}

\clearpage
\section*{Checklist}

\begin{enumerate}

\item For all authors...
\begin{enumerate}
  \item Do the main claims made in the abstract and introduction accurately reflect the paper's contributions and scope?
    \answerYes{}
  \item Did you describe the limitations of your work?
    \answerNo{Due to tight space constraints, we do not discuss it here, but we plan to submit a long-form journal paper in the near future with an extensive section discussing the limitations of the current work as well as potential for future work.}
  \item Did you discuss any potential negative societal impacts of your work?
    \answerNo{Our work specifically focuses on astronomical data and we do not see any obvious negative social impact of our work.}
  \item Have you read the ethics review guidelines and ensured that your paper conforms to them?
    \answerYes{}
\end{enumerate}

\item If you are including theoretical results...
\begin{enumerate}
  \item Did you state the full set of assumptions of all theoretical results?
    \answerNA{We do not include theoretical results.}
        \item Did you include complete proofs of all theoretical results?
    \answerNA{}
\end{enumerate}

\item If you ran experiments...
\begin{enumerate}
  \item Did you include the code, data, and instructions needed to reproduce the main experimental results (either in the supplemental material or as a URL)?
    \answerYes{We are preparing a github repository with data and code necessary to reproduce the results. This will be included in the accepted version.}
  \item Did you specify all the training details (e.g., data splits, hyperparameters, how they were chosen)?
    \answerYes{This is provided in Sections 2 and 3, with additional supplementary information in the Appendix.}
        \item Did you report error bars (e.g., with respect to the random seed after running experiments multiple times)?
    \answerNo{Due to tight space constraints, we do not discuss it here, but we have ran the same set of multiple seeds on both methods discussed in this paper and effects of variations across random seeds are not significant enough to change the main conclusions of this work.}
        \item Did you include the total amount of compute and the type of resources used (e.g., type of GPUs, internal cluster, or cloud provider)?
    \answerYes{This is included in the introduction of section 4.}
\end{enumerate}

\item If you are using existing assets (e.g., code, data, models) or curating/releasing new assets...
\begin{enumerate}
  \item If your work uses existing assets, did you cite the creators?
    \answerYes{Yes, in section 3, we cited the creators and provided footnote links to the relevant github pages of all libraries we used in this work.}
  \item Did you mention the license of the assets?
    \answerNo{Due to space constraints, we did not, but of the 3 libraries we used, \texttt{lenstronomy} uses a BSD 3-Clause "New" or "Revised" License, \texttt{deeplenstronomy} uses a MIT License, and \texttt{sbi} uses the GNU Affero General Public License v3.0.}
  \item Did you include any new assets either in the supplemental material or as a URL?
    \answerNo{}
  \item Did you discuss whether and how consent was obtained from people whose data you're using/curating?
    \answerNA{The libraries we used are publicly available.}
  \item Did you discuss whether the data you are using/curating contains personally identifiable information or offensive content?
    \answerNA{}
\end{enumerate}

\item If you used crowdsourcing or conducted research with human subjects...
\begin{enumerate}
  \item Did you include the full text of instructions given to participants and screenshots, if applicable?
    \answerNA{}
  \item Did you describe any potential participant risks, with links to Institutional Review Board (IRB) approvals, if applicable?
    \answerNA{}
  \item Did you include the estimated hourly wage paid to participants and the total amount spent on participant compensation?
    \answerNA{}
\end{enumerate}

\end{enumerate}

\end{document}